\documentclass{emulateapj}
\def\icarus{Icarus}
\usepackage{natbib}
\slugcomment{ApJ Letters, in press}

\begin{document}

\title{The surface composition of large Kuiper belt object 2007 OR10}
\author{M.E. Brown\altaffilmark{a}}
\author{A.J.\ Burgasser\altaffilmark{b,c,d}}
\author{W.C. Fraser\altaffilmark{a}}

\altaffiltext{a} {Division of Geological and Planetary Sciences, California Institute
of Technology, Pasadena, CA 91125}
\altaffiltext{b}{Center for Astrophysics and Space Science, University of California San Diego, La Jolla, CA 92093, USA; aburgasser@ucsd.edu}
\altaffiltext{c}{Massachusetts Institute of Technology, Kavli Institute for Astrophysics and Space Research, 77 Massachusetts Avenue, Cambridge, MA 02139, USA}
\altaffiltext{d}{Hellman Fellow}

\begin{abstract}
We present photometry and spectra of the large Kuiper belt object 2007 OR10.
The data show significant
near-infrared
absorption features due to water ice.
While most objects in the Kuiper belt with water
ice absorption this prominent have the optically neutral colors of water
ice, 2007 OR10 is among the reddest Kuiper belt objects known. One other
large Kuiper belt object -- Quaoar -- has similar red coloring and
water ice absorption, and it is hypothesized that the red coloration
of this object is due to irradiation of the small amounts of methane
able to be retained on Quaoar. 2007 OR10, though warmer than Quaoar,
 is in a similar volatile retention because it is sufficiently larger
that its stronger gravity can still retain methane. We propose, therefore, that
the red coloration on 2007 OR10 is also caused by the retention of
small amounts of methane. 
Positive detection will require spectra of methane on 2007 OR10 will
require spectra
with higher signal-to-noise. Models for volatile retention on Kuiper
belt objects appear to continue to do an excellent job reproducing all of the
available observations.
\end{abstract}


\section{Introduction}

The large majority of Kuiper belt objects (KBOs) contain no detectable
volatile ices on their surfaces, but a small number of the largest
objects have been found to have signatures of CH$_4$, CO, or N$_2$,
all ices with high vapor pressures at Kuiper belt temperatures.
After the discovery of volatiles on the surfaces
of Eris \citep{2005ApJ...635L..97B}, Makemake \citep{2007AJ....133..284B}, and Sedna \citep{2005A&A...439L...1B},
\citet{2007ApJ...659L..61S} proposed a simple method for assessing 
the possibility
of volatile retention on KBOs. For each relevant ice, they
compared the volatile loss due to Jean's escape -- the slowest
of many possible escape mechanisms -- to the total volatile inventory
of the object and divided the Kuiper belt into objects
which could and could not have retained that ice over the age of the
solar system. Only a handful of objects are massive enough
or cold enough to be able to retain volatiles.
Their model provided a compelling explanation of
the low abundance of N$_2$ on Makemake
\citep{2007AJ....133..284B,2008Icar..195..844T}, which is
smaller than Pluto and Eris,
and was also used to successfully predict
the presence of methane on Quaoar \citep{2007ApJ...670L..49S}.
To date, the volatile retention model has been completely successful
predicting which objects will and which will not have detectable surface
volatiles, with the unique exception being the large KBO Haumea,
which is the parent body of the only collisional family known in
the Kuiper belt \citep{2007Natur.446..294B} and clearly had an unusual 
history.

We provide an update to the \citet{2007ApJ...659L..61S} calculations in Figure 1.
We have used new vapor pressure data from \citet{2009P&SS...57.2053F} 
and, where possible,
have used measured sizes and masses of the largest KBOs. For Quaoar, the 
current measured diameter is 890 $\pm$ 70 km implying
a density of 4.2 $\pm$ 1.3 g cm$^{-3}$ \citep{2010ApJ...714.1547F}, 
but we assume the upper limit of size as the smaller sizes
lead to physically implausible densities. (Note that even
for the smaller size and higher density, however, Quaoar is
still expected to retain surface methane.) 
The size of 2007 OR10 is unmeasured, so, as will be justified 
below, we
assume that it has an albedo identical to the 0.18 albedo
of Quaoar that gives the size that we assume above, though
we allow albedo uncertainties of 50\% in either direction.
As in \citet{2007ApJ...659L..61S}, we calculate an "equivalent temperature" for each object by 
integrating the volatile loss through the object's entire orbit
and determining the temperature that an object in a circular
orbit would have to have to lose the volatile at that rate.
\begin{figure}
\plotone{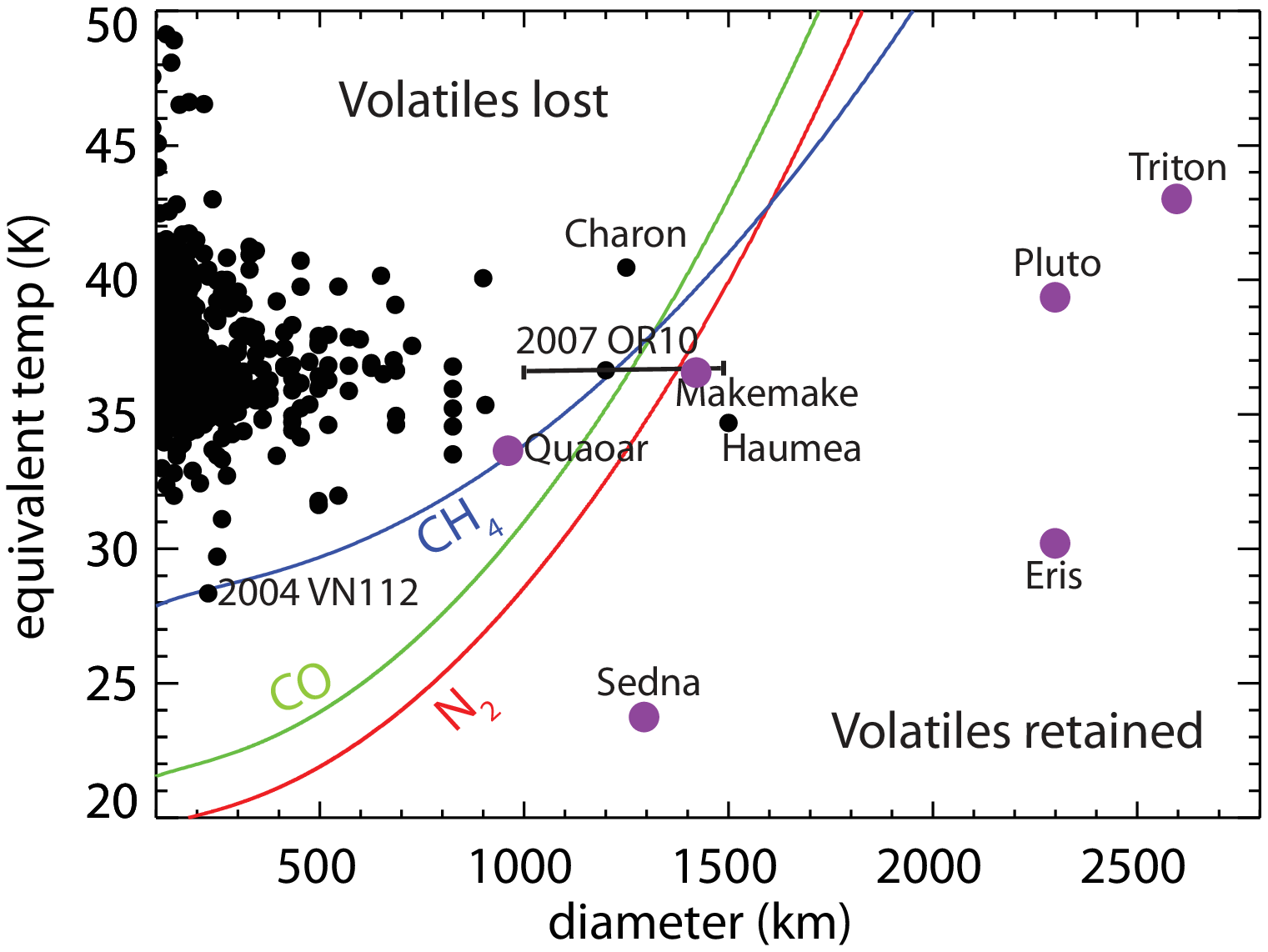}
\caption{A plot of volatile retention and loss in the Kuiper belt, updated
from \citet{2007ApJ...659L..61S}. Objects to the left of the CH$_4$, CO, and
N$_2$ lines are too small and too hot to retain any of those
surface volatiles
over the age of the solar system, while objects to the right can retain those
volatiles. All objects shown in purple have had CH$_4$ measured on their
surfaces. Some have additionally had N$_2$ or CO detected. No objects to 
the left of the lines have had any of these volatiles detected.}
\end{figure}

For our assumed albedo range, 2007 OR10 is somewhere between
the fourth and seventh largest
object known in the Kuiper belt. Its potential size
spans the range between the small volatile poor objects and the handful
of volatile rich objects. 2007 OR10 is thus an excellent test object 
for our understanding of volatile retention in the outer solar system.
We explore the surface composition of this object below using a 
combination of near-IR spectroscopy and multi-wavelength photometry.

\section{Observations}
The low-resolution, near-infrared spectrum of 2007 OR10 
was obtained on 2010 September 20 (UT) using the
Folded-port Infrared Echellette (FIRE) spectrograph 
on the 6.5m Magellan Baade Telescope
\citep{2008SPIE.7014E..27S,2010SPIE.7735E..38S}.  
FIRE's prism-dispersed mode provides continuous 
coverage of the 0.85--2.45~$\micron$ 
band with a variable resolution of 
$\lambda/\Delta\lambda$ = 250--350.
2007 OR10 was acquired and its motion confirmed
using FIRE's $J$-band imaging channel. The source was 
maintained on the 0$\farcs$6 slit 
by manual corrections to sidereal tracking.
Two series of ABBA dither exposure sequences were obtained 
with integrations of 120~s 
at an average airmass of 1.04.  These were 
followed by a single ABBA sequence of the
G2~V star HD~211544 ($V$=10.9) at a similar airmass.  
Exposures of a quartz flat field lamp (set at 1.2~V and 2.2~V)
and arc lamps (NeAr) were obtained for pixel response and wavelength calibration.
Data were reduced using the methods described in \citet{2007AJ....133..284B}.
The spectrum was converted into relative reflectance as well as 
corrected for telluric absorption and instrument response by dividing the
raw spectrum of 2007 OR10 by the spectrum of the solar type star HD~211544.

Photometry were obtained with the Wide-Field Camera 3 on the Hubble Space Telescope during cycles 17 (GO Program 11644)  and 18 (GO Program 12234). 
In cycle 17,  two 130 s exposures were taken in the F606W and F814W filters, 
and two 453 s exposures were taken in the F139M, and F153M filters. 
During the cycle 18 observations, two exposures were acquired 
in each of the F606W, F775W, F098M, F110W with exposure times of 
128s, 114s, 115s, and 207 s respectively. 
As well, four exposures of 275 s were acquired in the F127m filter. 
For both the cycle 17 and 18 observations, 3" dithers 
were applied between image pairs to reduce the effects of 
cosmic rays and pixel defects, with the exception of 
the F127m observations, in which 2 images were taken at each 
dither position. All observations in a cycle were acquired 
within a single orbit, minimizing the effect of any 
light curve OR10 may have. All data were processed 
through CalWFC3 version 2.3, the standard WFC3 image 
processing pipeline \citep{Rajan2010}. Circular 
apertures were used to measure the photometry. 
{\it Tiny Tim} version 7.1  
PSFs \citep{Krist1993}were used to generate infinite 
aperture corrections as well as interpolate over 
any bad pixels flagged during the image reductions. 

Fluxes were converted to relative reflectance by comparing to 
fluxes computed using the the {\it calcphot} routine
for a model solar spectrum 
\citep{Kurucz1979a} provided as part of the  {\it iraf} package {\it stsdas.synphot}. Approximate absolute reflectances were then obtained by
scaling the F606W relative reflectance to a value of 0.18, our assumed
albedo of 2007 OR10. All are shown in Table 1. Before calculating relative reflectances,
the Cycle 18 magnitudes were adjusted upward by 0.03 to account
for the difference in the F606W magnitudes between the two epochs. The
small magnitude difference is an expected consequence of object rotation.
\begin{deluxetable}{cccc}
\tablehead{\colhead{filter} & \colhead{2007 OR10} & \colhead{sun} & \colhead{reflectance} \\ 
\colhead{} & \colhead{(mag)} & \colhead{(mag)} & \colhead{} } 

\startdata
Cycle 17 \\
F606W &     21.68$\pm$ 0.02 &  -27.00 &  0.18 \\
F814W &     21.39$\pm$ 0.01 &  -26.54 &  0.37 \\
F139M &     22.06$\pm$ 0.01 &  -25.34 &  0.59 \\
F153M &     22.47$\pm$ 0.02 &  -25.09 &  0.51 \\
Cycle 18 \\
F606W &     21.65$\pm$0.02 &  -27.00 &  0.18 \\
F775W &     21.32$\pm$0.02 &  -26.64 &  0.34 \\
F098M &     21.37$\pm$0.01 &  -26.13 &  0.52 \\
F110W &     21.62$\pm$0.02 &  -25.78 &  0.58 \\
F127M &     21.84$\pm$0.01 &  -25.54 &  0.58 \\
\enddata

\end{deluxetable}

\section{Results}
Figure 2 shows the FIRE reflectance spectrum with the WFC3 
photometric points overlaid. We scale all point to an albedo
of 0.18 in the F606W filter, though we note that the true
value of the albedo has not been measured.
To increase the signal-to-noise in the FIRE spectrum, we also plot the median of 
every 32 spectral channels, oversampled by a factor of two, to simulate 
how the spectrum would appear to a lower
 resolution spectrograph. Uncertainties
on these data points are obtained by calculating the median 
absolute deviation of
eache 32 channel sample, 
which we multiply by 1.48 to obtain what would be the standard
deviation in a normally distributed sample, and then divide by the square root
of the number of spectral channels, approximating the standard deviation.
\begin{figure}
\plotone{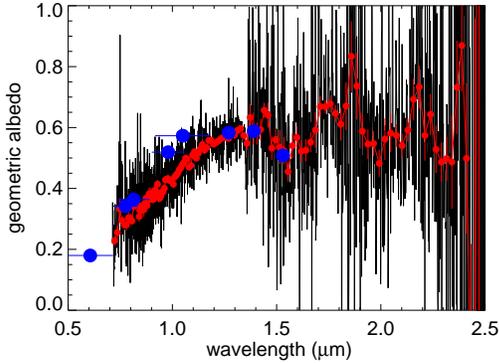}
\caption{The absolute reflectance spectrum of 2007 OR10. The gray lines 
show the full resolution FIRE spectrum of 2007 OR10. Regions of poor
atmospheric transmission are easily identified by the substantial increase in 
noise. To increase the signal-to-noise, we show a 32-channel median filtered
version of the spectrum in red, with error bars derived from the scatter
in the 32 channels. The large blue points show the WFC3 photometry, with
the horizontal error bars representing the width of each of the filters
used. The uncertainties in the WFC3 photometry are smaller than the size
of the data points. As the albedo of 2007 OR10 is not know, 
both data sets are arbitrary normalized. The FIRE and WFC3 data both clearly show a
$\sim$1.5$\mu$m absorption feature, while the FIRE spectrum also shows an
absorption around 2.0 $\mu$m. These absorption features are characteristic of
water ice on the surface.}
\end{figure}

The FIRE spectrum and the WFC3 photometry are in broad agreement 
in the area of overlap, 
though the match is imperfect. We suspect that the differences are due 
to differential refraction in the FIRE data, which, for these early
attempts at tracking a moving object, were not obtained with the slit aligned 
along the parallactic 
angle. 
Both data sets show, in particular, a very
red optical slope and a distinct absorption around 1.5 $\mu$m. The
FIRE spectrum shows an additional broad absorption feature near 2.0 $\mu$m
and, potentially, additional features redward.
Absorptions at 1.5 and 2.0 $\mu$m are the characteristics features of
water ice, which is frequently found on the largest KBOS \citep{2008AJ....135...55B}.

Figure 3 compares the spectrum of 2007 OR10 to a modeled spectrum of
a surface consisting of a mixture of water ice and a neutral material.
We place no special significance on the precise water ice surface model, as
many different types of specific parameters yield similar modeled spectra,
but, for concreteness we use the water ice absorption coefficients
of \citet{1998JGR...10325809G} and construct a simple 
Hapke model \citep{1993tres.book.....H} with 50$\mu$m grains
at 50 K spatially mixed with equal amounts of
a neutral material with an albedo of 80\%. The model spectrum is
sampled at the same resolution as the smoothed spectrum of the object.
Even at the low signal-to-noise ratio of the data, the water ice model 
provides an excellent match to the spectrum redward of 1.4 $\mu$m. 
At shorter wavelengths, however, 2007 OR10 is significantly redder than 
water ice. 
\begin{figure}
\plotone{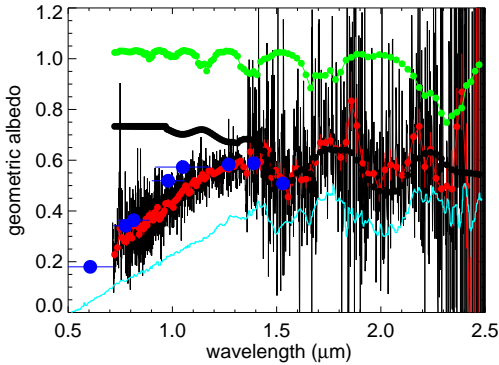}
\caption{The spectrum of 2007 OR10 compared to simple surface models. 
The black points show a modeled spectrum with a mixture of water ice and
a neutral material. This model clearly reproduces the prominent absorption
features at 1.5 and 2.0 $\mu$m, but cannot account for the very red
color of the object. Such a coloration on the equally-water-ice-rich object
Quaoar is hypothesized to be due to irradiation of remnant methane,
which is detected on Quaoar. 
The spectrum of Quaoar \citep{2004Natur.432..731J}
is shown for comparison in light blue, offset downward
by 0.15 units for clarity.
The green points show a modeled spectrum with a mixture of methane ice and a neutral material, shifted upward by
0.3 units for clarity. While the
signal-to-noise of the 2007 OR10 spectrum is insufficient to positively
detect any methane features, the large deviation from the water ice
model at 2.3$\mu$m occurs near the largest absorption expected from
methane. 
}
\end{figure}

Most of the KBOs with significant water ice absorption, such as Orcus, Haumea,
and the Haumea family members, have nearly neutral optical colors
\citep{2008AJ....135...55B}.
The most notable exception to this trend is Quaoar, which has
water ice absorptions nearly identical in depth to those of 2007 OR10 and
is almost as red. A spectrum of Quaoar \citep{2004Natur.432..731J}
over the same wavelength range
is shown, for comparison, in Figure 3. The coloring of Quaoar was suggested by
\citet{2007ApJ...670L..49S} to be due to the effects of methane, which turns
red with irradiation \citep{2006ApJ...644..646B}. Quaoar, as seen in Figure 1, is
barely large enough or cold enough to retain methane on its surface
and the amount left is small and detectable only at high signal-to-noise.

2007 OR10, depending on its precise size, could be in a similar 
regime of volatile retention as Quaoar.  If true, the Quaoar-like
optical color of 2007 OR10 could be a signature of the retention 
and irradiation of methane on the KBO. Indeed, if 2007 OR10 is assumed
to have the same albedo as Quaoar -- as the close match in visible and
near-infrared spectra might suggest -- 2007 OR10 sits in almost precisely
the same volatile loss regime as Quaoar. 
While the perihelion distance of 2007 OR10 of only 33.6 AU (compared to
Quaoar's of 41.6 AU) makes it significantly hotter than Quaoar, the
significantly larger size of 2007 OR10 allows its larger gravitational
pull to nonetheless potentially retain methane. 

If the hypothesis that the extreme red coloration on 2007 OR10 is --
like that of Quaoar -- caused 
by the irradiation of a small amount of remaining surface methane is correct,
we predict that methane absorption should be visible on
2007 OR10 as it is on Quaoar. Figure 3 also shows a simple model of a surface
including solid methane and a neutral component. The model is simply
the water model from above but with the absorption coefficients 
of methane \citep{2002Icar..155..486G} replacing those of water. The spectrum
is shifted upward by 0.3 units for clarity. 
 The strongest absorption
feature of methane, at 2.3 $\mu$m does indeed correspond with a large
variation from the water ice model. 
We conclude that while these data do not 
have sufficient signal-to-noise, particularly in the K-band, to
positively detect methane on 2007 OR10, the existence of this
volatile is plausible, and would provide a pleasing explanation
for the extreme red coloration of the KBO.

The large Kuiper belt object 2007 OR10 provides an excellent test of
our understanding of volatile loss and retention on the surfaces of
objects in the outer solar system. While the size of 2007 OR10 has yet
to be measured, the simple assumption that it has an identical 
albedo to Quaoar -- the object whose spectrum its spectrum most
resembles -- places 2007 OR10 into a regime where it would be expected
to retain trace amounts of methane on its surface. Such an object
would be expected to have red optical coloration from methane irradiation,
which both Quaoar and 2007 OR10 do have. In addition, such an object should 
have detectable signatures of methane if observed at sufficient
signal-to-noise. Such methane signatures have been detected on Quaoar, 
but require higher signal-to-noise to positively identify on 2007 OR10.
While additional measurements of the size and spectrum of
2007 OR10 are clearly required, we conclude that volatile retention
models \citep{2007ApJ...659L..61S} appear to continue to flawlessly
predict both the presence and absence of volatiles on all objects
in the Kuiper belt which have been observed to date.



\begin{thebibliography}{20}
\expandafter\ifx\csname natexlab\endcsname\relax\def\natexlab#1{#1}\fi

\bibitem[{{Barkume} {et~al.}(2008){Barkume}, {Brown}, \&
  {Schaller}}]{2008AJ....135...55B}
{Barkume}, K.~M., {Brown}, M.~E., \& {Schaller}, E.~L. 2008, \aj, 135, 55

\bibitem[{{Barucci} {et~al.}(2005){Barucci}, {Cruikshank}, {Dotto}, {Merlin},
  {Poulet}, {Dalle Ore}, {Fornasier}, \& {de Bergh}}]{2005A&A...439L...1B}
{Barucci}, M.~A., {Cruikshank}, D.~P., {Dotto}, E., {Merlin}, F., {Poulet}, F.,
  {Dalle Ore}, C., {Fornasier}, S., \& {de Bergh}, C. 2005, \aap, 439, L1

\bibitem[{{Brown} {et~al.}(2007{\natexlab{a}}){Brown}, {Barkume}, {Blake},
  {Schaller}, {Rabinowitz}, {Roe}, \& {Trujillo}}]{2007AJ....133..284B}
{Brown}, M.~E., {Barkume}, K.~M., {Blake}, G.~A., {Schaller}, E.~L.,
  {Rabinowitz}, D.~L., {Roe}, H.~G., \& {Trujillo}, C.~A. 2007{\natexlab{a}},
  \aj, 133, 284

\bibitem[{{Brown} {et~al.}(2007{\natexlab{b}}){Brown}, {Barkume}, {Ragozzine},
  \& {Schaller}}]{2007Natur.446..294B}
{Brown}, M.~E., {Barkume}, K.~M., {Ragozzine}, D., \& {Schaller}, E.~L.
  2007{\natexlab{b}}, \nat, 446, 294

\bibitem[{{Brown} {et~al.}(2005){Brown}, {Trujillo}, \&
  {Rabinowitz}}]{2005ApJ...635L..97B}
{Brown}, M.~E., {Trujillo}, C.~A., \& {Rabinowitz}, D.~L. 2005, \apjl, 635, L97

\bibitem[{{Brunetto} {et~al.}(2006){Brunetto}, {Barucci}, {Dotto}, \&
  {Strazzulla}}]{2006ApJ...644..646B}
{Brunetto}, R., {Barucci}, M.~A., {Dotto}, E., \& {Strazzulla}, G. 2006, \apj,
  644, 646

\bibitem[{{Fraser} \& {Brown}(2010)}]{2010ApJ...714.1547F}
{Fraser}, W.~C. \& {Brown}, M.~E. 2010, \apj, 714, 1547

\bibitem[{{Fray} \& {Schmitt}(2009)}]{2009P&SS...57.2053F}
{Fray}, N. \& {Schmitt}, B. 2009, \planss, 57, 2053

\bibitem[{{Grundy} \& {Schmitt}(1998)}]{1998JGR...10325809G}
{Grundy}, W.~M. \& {Schmitt}, B. 1998, \jgr, 103, 25809

\bibitem[{{Grundy} {et~al.}(2002){Grundy}, {Schmitt}, \&
  {Quirico}}]{2002Icar..155..486G}
{Grundy}, W.~M., {Schmitt}, B., \& {Quirico}, E. 2002, Icarus, 155, 486

\bibitem[{{Hapke}(1993)}]{1993tres.book.....H}
{Hapke}, B. 1993, {Theory of reflectance and emittance spectroscopy} (Topics in
  Remote Sensing, Cambridge, UK: Cambridge University Press, |c1993)

\bibitem[{{Jewitt} \& {Luu}(2004)}]{2004Natur.432..731J}
{Jewitt}, D.~C. \& {Luu}, J. 2004, \nat, 432, 731

\bibitem[{{Krist}(1993)}]{Krist1993}
{Krist}, J. 1993, in Astronomical Society of the Pacific Conference Series,
  Vol.~52, Astronomical Data Analysis Software and Systems II, ed. R.~J.
  {Hanisch}, R.~J.~V. {Brissenden}, \& J.~{Barnes}, 536--+

\bibitem[{{Kurucz}(1979)}]{Kurucz1979a}
{Kurucz}, R.~L. 1979, \apjs, 40, 1

\bibitem[{{Rajan} {et~al.}(2010){Rajan}, {Quijano}, {Bushouse}, \&
  {Deustua}}]{Rajan2010}
{Rajan}, A., {Quijano}, J.~K., {Bushouse}, H., \& {Deustua}, S. 2010, {WFC3
  Data Handbook V2} (StSci:Baltimore)

\bibitem[{{Schaller} \& {Brown}(2007{\natexlab{a}})}]{2007ApJ...670L..49S}
{Schaller}, E.~L. \& {Brown}, M.~E. 2007{\natexlab{a}}, \apjl, 670, L49

\bibitem[{{Schaller} \& {Brown}(2007{\natexlab{b}})}]{2007ApJ...659L..61S}
---. 2007{\natexlab{b}}, \apjl, 659, L61

\bibitem[{{Simcoe} {et~al.}(2008){Simcoe}, {Burgasser}, {Bernstein}, {Bigelow},
  {Fishner}, {Forrest}, {McMurtry}, {Pipher}, {Schechter}, \&
  {Smith}}]{2008SPIE.7014E..27S}
{Simcoe}, R.~A., {Burgasser}, A.~J., {Bernstein}, R.~A., {Bigelow}, B.~C.,
  {Fishner}, J., {Forrest}, W.~J., {McMurtry}, C., {Pipher}, J.~L.,
  {Schechter}, P.~L., \& {Smith}, M. 2008, in Society of Photo-Optical
  Instrumentation Engineers (SPIE) Conference Series, Vol. 7014, Society of
  Photo-Optical Instrumentation Engineers (SPIE) Conference Series

\bibitem[{{Simcoe} {et~al.}(2010){Simcoe}, {Burgasser}, {Bochanski},
  {Schechter}, {Bernstein}, {Bigelow}, {Pipher}, {Forrest}, {McMurtry},
  {Smith}, \& {Fishner}}]{2010SPIE.7735E..38S}
{Simcoe}, R.~A., {Burgasser}, A.~J., {Bochanski}, J.~J., {Schechter}, P.~L.,
  {Bernstein}, R.~A., {Bigelow}, B.~C., {Pipher}, J.~L., {Forrest}, W.,
  {McMurtry}, C., {Smith}, M.~J., \& {Fishner}, J. 2010, in Society of
  Photo-Optical Instrumentation Engineers (SPIE) Conference Series, Vol. 7735,
  Society of Photo-Optical Instrumentation Engineers (SPIE) Conference Series

\bibitem[{{Tegler} {et~al.}(2008){Tegler}, {Grundy}, {Vilas}, {Romanishin},
  {Cornelison}, \& {Consolmagno}}]{2008Icar..195..844T}
{Tegler}, S.~C., {Grundy}, W.~M., {Vilas}, F., {Romanishin}, W., {Cornelison},
  D.~M., \& {Consolmagno}, G.~J. 2008, \icarus, 195, 844

\end{thebibliography}
\end{document}